# Small Spacecraft for Global Greenhouse Gas Monitoring

Victoria Mayorova[a], Andrey Morozov[a], Iliya Golyak[a], Nikita Lazarev[a], Valeriia Melnikova[a,*], Dmitry Rachkin[a], Victor Svirin[a], Stepan Tenenbaum[a], Igor Fufurin[a]

[a] Bauman Moscow State Technical University, 105005, Russia, Moscow, 2nd Baumaskay st., 5-1

*Corresponding author. E-mail: melnikovabg@bmstu.ru, Tel: +79105831865

**ABSTRACT** This work is devoted to the capabilities analysis of constellation and small spacecraft developed using CubeSat technology to solve promising problems of the Earth remote sensing in the area of greenhouse gases emissions. This paper presents the scientific needs for such tasks, followed by descriptions and discussions of the micro-technology application both in the small satellite platform design and in the payload design. The overview of analogical spacecraft is carried out. The design of a new spacecraft for determination the oxygen and carbon dioxide concentration in the air column along the line of sight of the spacecraft when it illuminated by reflected sunlight is introduced. A mock-up of the device was made for greenhouse gases remote sensing a Fourier Transform Infrared (FTIR) spectroradiometer is placed in the small spacecraft design. The results of long-term measurements of greenhouse gas concentrations using the developed Fourier spectrometer mock-up is presented.

**KEYWORDS** CubeSat, remote sensing, greenhouse gas, emissions, carbon, environmental footprint, air pollution monitoring.

## 1. Introduction

Currently, an urgent task is to create a carbon calculator that allows assessing the ecological footprint of technologies. Anthropogenic greenhouse gas emissions come from a very large number of point sources (including oil and gas facilities, coal mines, landfills, sewage treatment plants and animal husbandry). Ground detection technologies currently in use are capable of capturing gases at low detection thresholds, but they are limited in terms of application time and territorial coverage. Satellite monitoring can contribute to the rapid and fairly accurate identification of problematic greenhouse gas emissions. Restrictions on the observation time from satellites can be resolved by launching more satellites in a constellation (by creating a grouping).

Consumers of such information can be: the oil and gas enterprises (monitoring of infrastructure facilities for incidents with $CH_4$ emissions – to optimize processes and reduce economical losses), The Federal Services for Supervision of Natural Resources, the Ministries of Natural Resources (monitoring of incidents with $CO_2$ emissions as a pollutant – assessment and justification of fines required to be paid to the budget), the Government (for example control of the implementation plan Strategies of socio-economic development of the Russian Federation with a low level of greenhouse gas emissions until 2050 – assessment of indicators on greenhouse gas emissions).

This work is devoted to the analysis of the capabilities of constellation and small spacecraft developed using CubeSat technology to solve promising problems of the Earth remote sensing in the area of greenhouse gas emissions.

## 2. Development State Overview

Hydrometeorological spacecraft, such as Meteor-M [1], Metop [2], Terra/Aqua, Sentinel-5P [3], which are equipped with infrared spectrometric and multispectral/hyperspectral equipment are successfully used to solve climate monitoring tasks. Such equipment allows assessing the atmosphere chemical composition (spatial resolution of 10-100 km, viewing bandwidth up to 2500 km) and the ecosystems state based on the surface spectral image. There are also specialized satellites, such as GOSAT, OCO 1, 2 [4], equipped with infrared spectrometers for more accurate localization in determining the content of greenhouse gases in the Earth's atmosphere than the equipment of hydrometeorological devices (spatial resolution of 1-10 km, viewing bandwidth up to 100 km). These devices are distinguished by high values of technical characteristics, productivity and informative value of the data obtained, however, due to their small number, they are able to solve the tasks of operational monitoring (several days) and search for point sources of atmospheric pollution to a limited extent, the increase in the number of satellites is stopped by budgetary constraints of space projects, therefore, instrument developers are forced to look for the optimum between the orbit parameters, namely the ability to monitor the same surface areas – the multiplicity of the sun-synchronous orbit and the spatial and spectral resolution determined by the viewing band width and the detectors sensitive elements technical characteristics. Since 2016, a project GHGSat (Canada) [5, 6] has been developing for a small satellite constellation. Small spacecraft equipped with an imaging spectrometer for measuring methane content with a spatial resolution of 50 m and a viewing bandwidth of 10 km. The developers have launched 6 of the 12 planned spacecraft into orbit. The data obtained from GHGSat devices have already demonstrated their significance and relevance in terms of searching for pollution point sources. In particular, the system captured the release of methane that occurred on January 14, 2022 at the Raspadskaya mine, Kemerovo region [7] and in September 2022 photographed the site of a methane leak from the Nord Stream gas pipeline [8].

Nowadays, within the framework of the Federal Space Program in the Russian Federation, the Meteor-M №2 [9] spacecraft has been created, launched into orbit and is being used for its intended purpose - climate monitoring. Four more Meteor-M spacecraft are planned to be launched until 2025. These satellites are equipped with an infrared Fourier spectrometer, which allows measurements of the gas composition of the atmosphere with a 30 km view field and a maximum bandwidth of up to 2500 km (with a point step of 100 km), which is not enough for gases point sources operational monitoring.

Based on the analysis results, it can be concluded that many scientific equipment have large dimensions and cannot be installed in a CubeSat, therefore, three types of devices are accepted for consideration to install in CubeSat: The Fabry-Perot spectrometer (SWIR), a heterodyne spectroradiometer and a Fourier Transform Infrared (FTIR) Spectroradiometer.

### 3. Spacecraft Design

The external and internal layout of the developed small spacecraft is made in the logic of the CubeSats standard [10]: the spacecraft has the form of a parallelepiped (Fig. 1). The volume of the spacecraft is 16U, the cross section of the spacecraft corresponds to the CubeSat 12U standard size: 226 × 226 mm defined in [10]. Four parallel ribs form the constitute rails for the spacecraft to exit the transport and launch pad. The satellite is divided into two parts: The Fourier Spectroradiometer unit and the service systems unit, created and assembled independently. Each of the compartments is installed to two detachable frames through special connecting frames. The main technical characteristics of the device are given in Table 1.

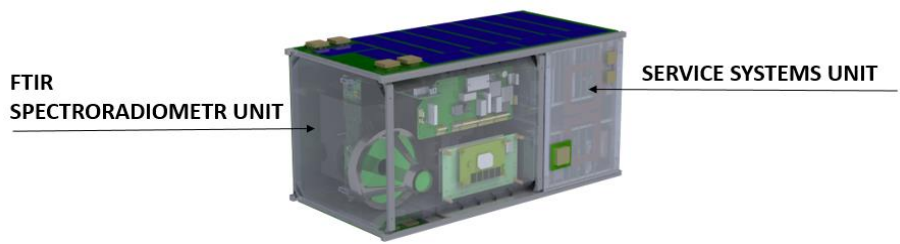

**Fig. 1.** Spacecraft internal design

**Table 1**
Spacecraft technical characteristics.

| Parameter | Value |
| --- | --- |
| Lifetime | > 3 years |
| Dimensions | 226x226x454 mm (CubeSat 16U) |
| Weight | 23 kg |
| Orbit parameters | from 500 to 600 km, SSO |
| Daily power (average orbital) | 10 W |
| Data transmission rates over radio channels: | |
|     - S (half duplex) | up to 4 Mbit/s |
|     - UHF (half duplex) | 9600 bps |
|     - X (Spacecraft-Ground) | 1 Gbit/s |
| Data storage | 32 Gb |
| Altitude control system | Three-axis: flywheels with unloading by magnetic coils |
| The orientation error (3σ) on all axes | < 0.4 deg |
| The stabilization error (3σ) on all axes | < 0.01 deg/sec |
| Propulsion system | electric ablative pulse |
| Payload | FTIR spectroradiometer unit |

## 3.1. FTIR Spectroradiometer Unit

FTIR spectroradiometer unit is integrated into the design of developed small spacecraft for greenhouse gases remote sensing. Namely, to determine the oxygen and carbon dioxide concentration in the air column along the line of sight of the device when it illuminated by reflected sunlight (Fig. 2). Carbon dioxide molecules are conventionally depicted in the atmosphere column, the content of which is detected the spacecraft. The FTIR spectroradiometer unit is made by the Center for Applied Physics of Bauman Moscow State Technical University [11]. This is a space version of their ground-based FTIR spectrometers.

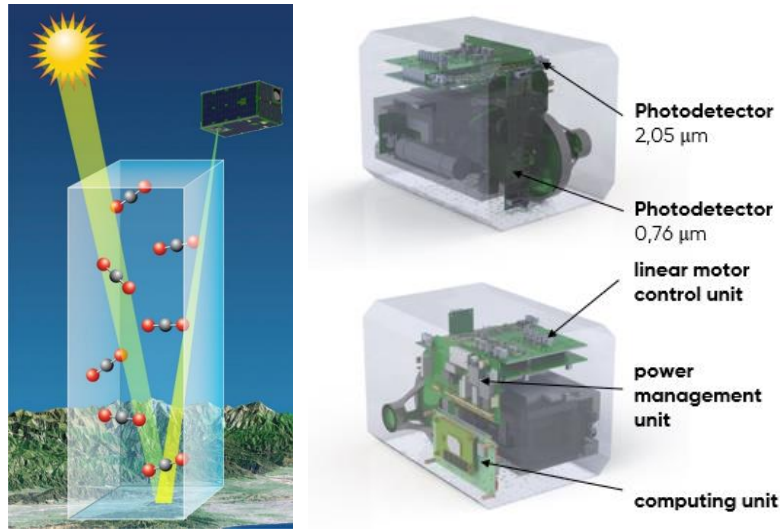

**Fig. 2.** The Fourier Spectroradiometer principle of operation and unit design

The device is based on an Michelson interferometer (Fig. 2). The satellite and scientific instruments should provide a target image and spectral information about the content of target gases: the first channel $O_2$ and the second channel $CO_2$. A distinctive feature compared to other missions is the combination of high spatial resolution and high spectral resolution in a compact volume. FTIR spectroradiometer optical scheme is shown on fig. 3.

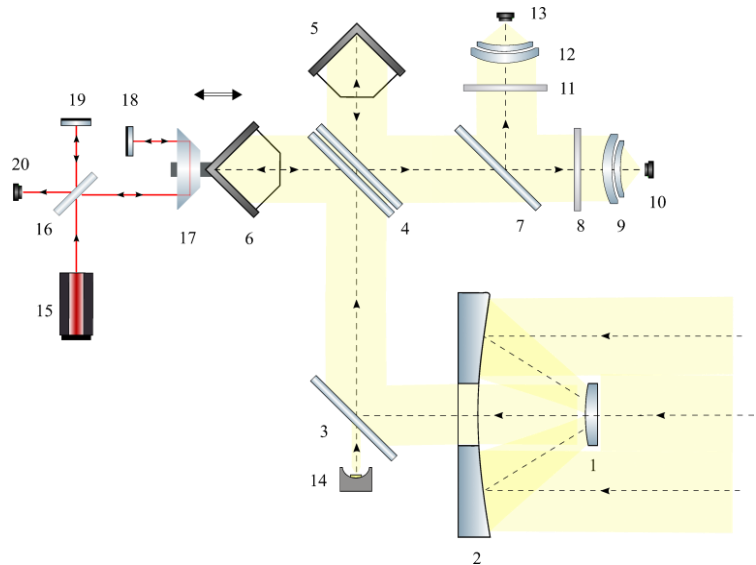

**Fig. 3.** The FTIR Spectroradiometer optical scheme

1, 2 – mirror lens, 3 – mirror, 4 – beam splitter, 5, 6 – corner reflectors, 7 – dichroic filter, 8 – bandpass filter with a length of 2.0-2.2 µm, 9 – focusing lens, 10 – radiation receiver 2.0 – 2.2 µm, 11 – bandpass filter for a length of 0.75-0.80 µm, 12– focusing lens, 13 – photodetector 0.75 - 0.80 µm, pos. 14 – reference radiation source, 15 – 632 nm laser, 16 – reference channel beam splitter, 17 – dihedron, 18, 19 – flat mirrors, 20 – reference channel photodetector

Tetrahedral reflectors are used in the receiving channel of the spectrometer to ensure the stability and reliability of its operation (pos. 5, 6) [12, 13] as mirrors with an aperture of 36 mm and a deviation of 1 angular second. The movable reflector is mounted on a spring parallelogram, which provides a stroke of 4 mm in one direction. The course of the movable mirror allows to register interferograms with the number of samples of 32 000, which determines the theoretical spectral resolution of 2 cm$^{-1}$ [14]. The device uses a beam splitter made of ZnSe (pos. 4) material and designed for the visible and infrared regions of the spectrum. The dichroic filter (pos. 7) is designed for spectral separation of the recorded radiation into visible and long-wave IR. Carbon dioxide ($CO_2$) absorption is recorded by an InGaAs receiver (pos. 10) with an active area of 1 mm and a detection capacity $D^*$ equal to $2.0 \cdot 10^{12}$ cm $\sqrt{Hz}$/W. A bandpass filter (pos. 8) is installed in front of the receiver for a spectral range of 2.0 – 2.2 μm.

Since the oxygen concentration in the lower atmosphere is permanent, the absolute concentration of gases can be calculated based on $O_2$ absorption line. Registration of oxygen ($O_2$) absorption lines is carried out by a Si photodetector (pos. 13) with an active area of 1 mm. The noise equivalent power (NEP) value for the detector is $6.2 \cdot 10^{-15}$ W/$\sqrt{Hz}$. To cut off visible radiation, a bandpass filter (pos. 11) is also installed in front of the receiver for a spectral range of 0.75 – 0.80 μm.

To stabilize the speed of the mirror and determine the moments of reading the interferogram, a reference channel with a fourfold difference in the optical path is used, which further leads to an increase in the accuracy of the sampling frequency. This is achieved by using an additional system consisting of a dihedron (17) and a flat mirror (18) [13]. A He-Ne laser with a wavelength of 632 nm is used as a reference source. To take into account the hardware function of the spectrometer [15, 16] and adjust the recorded IR absorption spectra, a reference radiation source (pos. 14) is installed.

Described FTIR spectroradiometer can be used for trace gas analysis [17] and remote sensing [18]. Special mathematical methods can be used [19, 20] for spectral analysis for detection of several gases and estimation its concentrations. The main technical characteristics of the spectroradiometer are given in Table 2.

**Table 2**

FTIR Spectroradiometer characteristics.

| Parameter | Value |
|---|---|
| Spectral range | |
| - First channel ($O_2$ line) | 0.75-0.80 μm |
| - Second channel ($CO_2$ line) | 2.0-2.2 μm |
| Spectral resolution | 2 cm$^{-1}$ |
| Angular field of view | 0.01 rad |
| Entrance aperture | 100 mm |
| Multiplicity of the input system | 4 |
| Diameter of the observation object | 5 km |
| Power consumption | < 100 W |
| Power supply | 12 V |
| Dimensions | 268 x 208 x 216 mm |
| Weight | < 10 kg |

The placement of the FTIR spectrometer on board the small spacecraft will allow constant monitoring of the concentration of a number of gases.

*3.2 Service Systems Unit*

Service system boards are installed in the service system unit on four racks. The preliminary sequence of installing the boards in each rack is shown in Fig. 3. Racks are installed on a large control board. The service systems were developed taking into account the flight tests of the «Yareelo»[1] №1 and 2 spacecraft [21-24].

---

[1] Slavic sun god

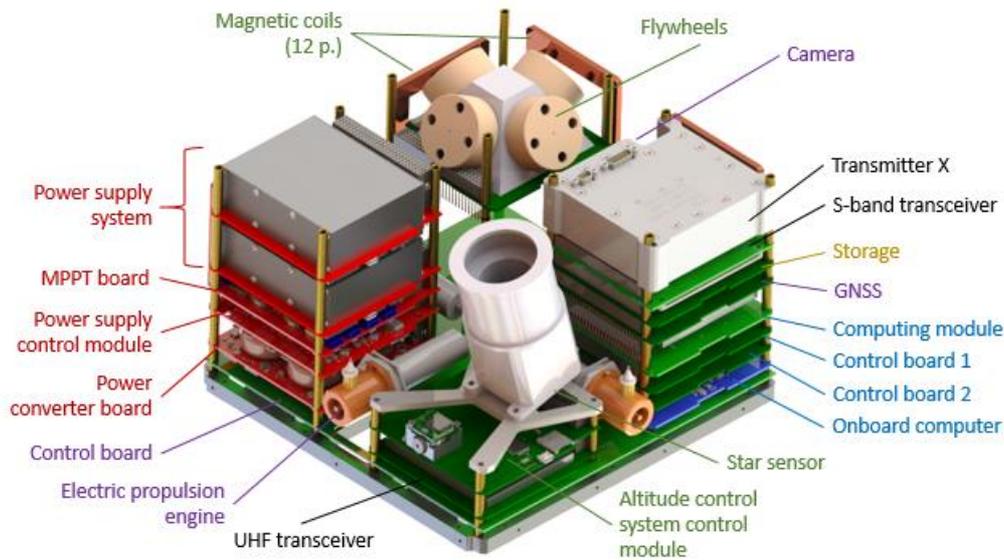

**Fig. 3.** Racks of boards with service systems

Due to the requirements orientation of the photodetector in nadir, the satellite has active three-axis altitude control system: flywheels [25] with unloading by magnetic coils. The magnetic coils are located 2 pieces on each of the external panels on the X and Y axes, and 4 pieces on the Z axis on the control board. To determine the orientation of the spacecraft, the data of the star sensor, solar sensor, accelerometer and magnetometer are used.

Power supply system provides power of 80 W and voltage of 12 V to the science equipment. Batteries are LiFePO4 elements of the ANR26650 type. These elements are characterized by large resource and high current output. Solar panels are built on the basis of AsGa solar cells with an efficiency of 28.8 % [26].

For transferring the target information to the ground control center the radio communication systems in UHF, S and X bands are used.

Three electric propulsion engines are used to change the satellite's orbit [27-29]. It is proposed to use teflon as a working medium, since it has a high density (2.21 g/cm$^3$) and low after–evaporation - a factor that reduces the characteristics of ablative pulsed plasma engine.

## 4. Spacecraft mock-up

A spacecraft mock-up was created (see fig. 4). External panels (including the installation of solar cells and antennas), the frames, the mock-up of the star sensor, the block of service systems (including the boards of systems and devices), the propulsion system and camera were manufactured.

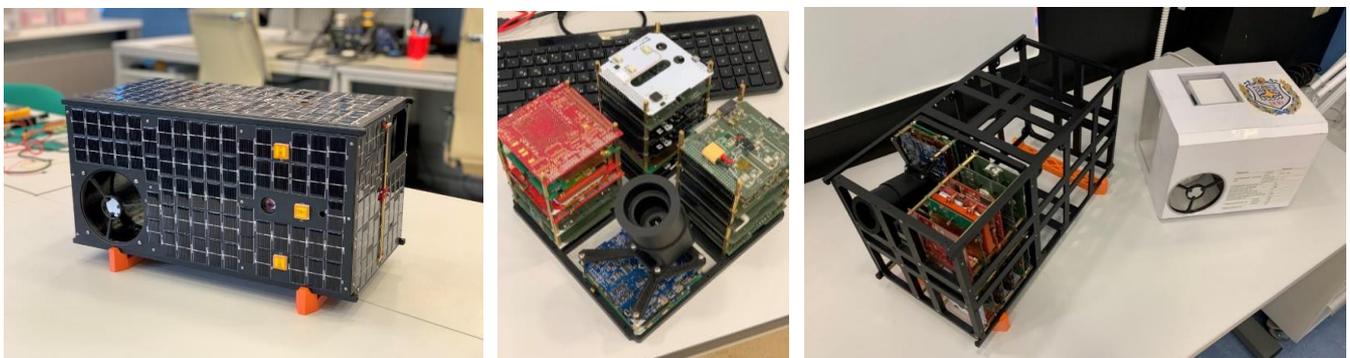

**Fig. 4.** Spacecraft mock-up design

## 5. Results and Discussion

We have developed a mock-up of the described satellite presented in current paper. Satellite's weight is about 23 kg, dimensions are 23×23×46 cm and 43% of the mass is taken up by the payload. Designed with distribution and manufacturability in mind, the satellite is optimized for global greenhouse gas monitoring tasks by using readily available parts (e.g., industrial electronic components) and inexpensive production conditions.

The optical scheme shown in Fig. 3 was implemented in the FTIR spectrometer mock-up sample. Spectral transmittance of the atmosphere obtained on the FTIR spectrometer is shown in Fig. 5. The spectrum was obtained by averaging over 15 interferograms with a total registration time of 1 min. The experimental spectrum contains oxygen absorption lines of $O_2$ at the wave number 7880 $cm^{-1}$, $CO_2$ at the wave numbers 6250 $cm^{-1}$ and 6350 $cm^{-1}$ and $CH_4$ at 6024 $cm^{-1}$. The position of the absorption lines corresponds to the one given in the HITRAN spectral base [30]. For single interferogram, the signal-to-noise ratio in the spectrum equals to 1220.

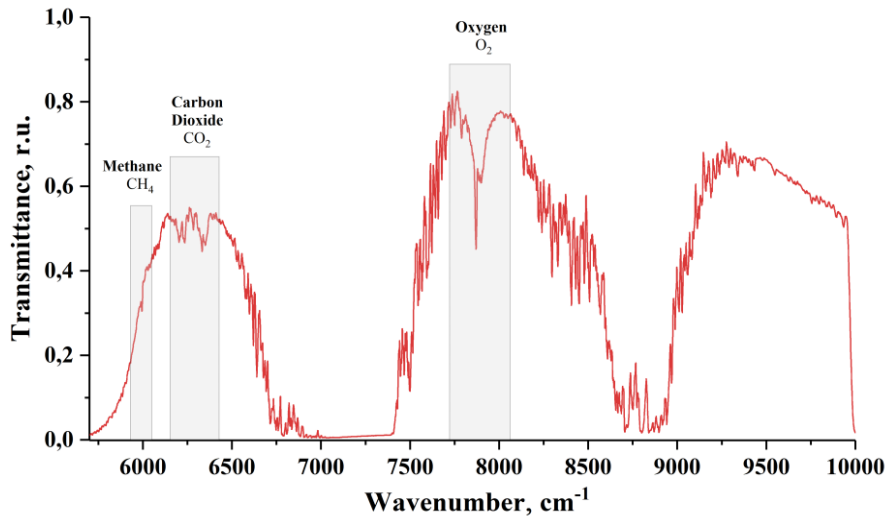

**Fig. 5.** The atmosphere transmittance spectrum measured by designed FTIR spectroradiometer

A long-term registration of greenhouse gases was carried out for 7 hours and their atmospheric content was calculated. Both integral and volumetric concentrations were calculated. Integral values of $CO_2$ and $CH_4$ were calculated from the depth of absorption lines in the recorded spectrum. The volume concentration of $CO_2$ and $CH_4$ was calculated from their integral concentrations, taking into account the normalization for the intensity of the $O_2$ absorption line.

Graphs of volume concentrations of $CO_2$, $CH_4$ are shown in Fig. 6. The shape of the graph represents the traffic jam in the observation region. In the evening, the decrease in volume concentration is associated with an increase in the optical path and the capture of air mass located outside Moscow.

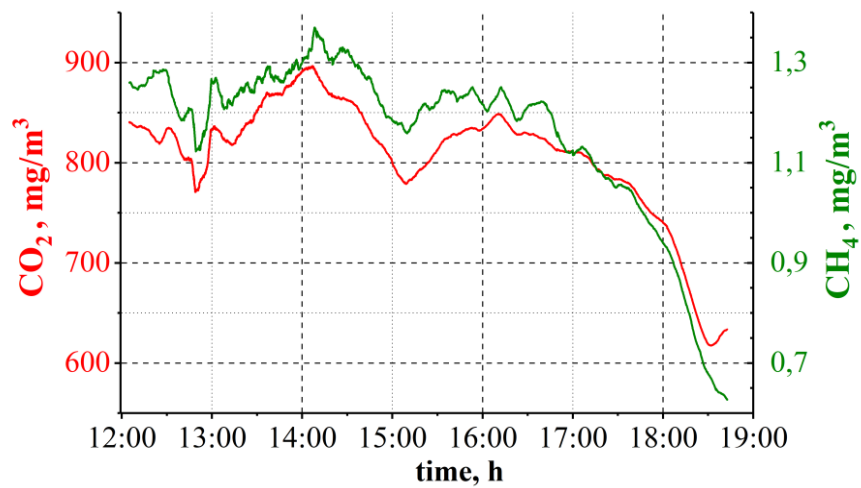

**Fig. 6.** A long-time registration of some greenhouse gases concentrations

## 6. Conclusions

The tools used to assess the technological footprint should include technologies with different capabilities that can be applied in different territorial and temporal coverage. It is incredibly important to conduct comprehensive operational observations of greenhouse emissions from various platforms — space, ground-based and over water. It must be a very complex, interdependent system that requires a comprehensive observing system.

The global constellation will require satellites and sensors in both low-Earth and geostationary orbits designed for long-term operation, in addition to ground-based sensors for tuning space instruments, verifying satellite data and creating complex atmospheric models. Currently, it seems relevant to create a grouping of small CubeSat-type spacecraft for monitoring point sources of greenhouse gas emissions with the possibility of a global overview to help large spacecraft. By continuously monitoring and pinpointing point source emissions at individual facilities, the microsatellite is designed to complement other observing systems capable of tracking net regional emissions and extremely large point sources. This surveillance strategy can be implemented by instructing satellites to collect data on priority areas based on a combination of prior knowledge of the location of the infrastructure and follow-up based on "tips" from other satellites designed to monitor a wider area. The presence of several satellites in orbit means that if one of them fails, the entire mission will not be compromised, coverage will decrease slightly, but it will still be possible to observe and collect data.

### Acknowledgments

The project was carried out within the framework of the «BaumanGoGreen» strategic project of the «Priority-2030» program of The Ministry of Science and Higher Education of the Russian Federation.

### Compliance with ethics guidelines

All authors declare that they have no conflict of interest or financial conflicts to disclose.